**PAPER • OPEN ACCESS**

# Randomized C/C++ dynamic memory allocator

View the article online for updates and enhancements.

# Randomized C/C++ dynamic memory allocator

**Irina Aleksandrovna Astrakhantseva**[1][0000-0003-2841-8639], **Roman Gennadevich Astrakhantsev**[2][0000-0001-9880-2826] **and Arseny Viktorovich Mitin**[2][0000-0001-8307-7625]

[1] Ivanovo State University of Chemistry and Technology, 7, Sheremetevskiy avenue, Ivanovo,153000, Russia

[2] National Research University Higher School of Economics, 20, Myasnitskaya str., Moscow, 101000, Russia

E-mail i.astrakhantseva@mail.ru

**Abstract**. Dynamic memory management requires special attention in programming. It should be fast and secure at the same time. This paper proposes a new randomized dynamic memory management algorithm designed to meet these requirements. Randomization is a key feature intended to protect applications from "use-after-free" or similar attacks. At the same time, the state in the algorithm consists only of one pointer, so it does not consume extra memory for itself. However, our algorithm is not a universal solution. It does not solve the memory fragmentation problem and it needs further development and testing.

## 1. Introduction
Most modern applications deal with large amounts of data, which size cannot be determined at compile time, so the memory should be managed dynamically in runtime. Dynamic memory management is one of the major challenges in programming since it should be time and memory efficient and secure. Time efficiency is often achieved by indexing free memory, but it is not memory efficient since indexes require memory space to be stored. Memory efficient algorithms usually lack metadata to optimize allocation time. This study endeavors to propose a randomized dynamic memory management algorithm, which tries to be both secure and efficient at the same time.

The rest of the paper is organized as follows. In Section 2, we demonstrate an overview of some existing ideas and techniques in this area. Section 3 contains an explanation of the proposed algorithm for memory management.

## 2. Materials and methods
Dynamic memory management is used in almost every complex software and requires special attention to make the allocation efficient and secure. In the following we present an overview of related work in this area.

One of the main security issues of dynamic memory allocation, the "use-after-free" problem was formulated by David Dewey et al. [1]. The paper focuses on dynamic memory management in the standard C++ library, which is unable to determine the new location of previously moved pointers. The same can be achieved in most compiled languages.

In addition, Qiang Zeng et al. [2] classifies different types of attachments and provides a way to patch programs to make them resistant against those attacks. The main disadvantage of this approach is that it is not memory-efficient since it requires additional 64 bits of metadata prepended before each allocated block and additional 4Kb for guard memory pages.

The alternative approach was suggested by Jonathan Ganz et al. [3]: to make the allocator return random address pointers. Most kernels of modern operating systems already use this approach internally since TSLF methods are one of the most efficient in terms of time complexity [4]. In contrast to the aforementioned methods, this one does not use hash maps and therefore is less memory consuming.

Another important dynamic memory management concern is fragmentation. Some programming languages with built-in garbage collectors, Java, for instance, takes an assumption that dynamically allocated objects act like "survivors" in memory space [5]: the longer the object exists, the less probability of being freed during the next GC iteration. Unfortunately, this only applies to languages with GC's since we are not able to change addresses of pointers at runtime in languages with manual dynamic memory management.

Another approach suggested by Nikola Zlatonov [6] is to divide dynamic memory space into pools for objects with corresponding block sizes in geometric progression. But it tends to be less memory efficient. Moreover, it applies only to specific types of applications [7].

Beichen Liu et al. (2019) [8] introduced the "SlimGuard" memory allocation approach, which is designed to be secure and time and memory efficient at the same time. We borrowed some techniques to use them in our allocator. In particular, we also prepend a random-sized free block before the block, which is being allocated.

As can be seen from the literature overview above, modern dynamic memory management approaches are aimed to be either time/memory efficient or secure but not both at the same time. Only experimental approaches try to find a balance between efficiency and security. In our study we provide a ready-to-use dynamic memory allocator which meets both efficiency and security requirements.

## 3. Results

*A. OS Memory Management*

Most modern operating system kernels provide a simple interface (syscalls) for applications in user space to request dynamic memory of a given size. The OS only maps memory pages inside the virtual address space of the calling process, so the application itself is responsible for further management and freeing allocated memory.

From the security perspective, if the allocation algorithm is fully determined at compile time, the hacker is able to decompile the executable file and predict the location of a particular object in memory of a process. Then, with high OS privileges, it is possible to attach to a process, read its memory and extract desired values, such as passwords, license keys, private keys, etc.

Modern operating systems already use address randomization, but only for addresses of memory pages, not the objects inside the application since the OS does not know anything about the application pointers. Protection provided by the operating system is not sufficient to ensure security level of application in terms of dynamic memory allocation. So, in-process dynamic memory management is required to make it secure.

*B. Basic Allocation*

Each allocated memory block is prepended with a so-called MemoryControlBlock (MCB) right before it. It holds metadata about the block and bidirectional linked list of free blocks. In C++ MemoryControlBlock has the following definition:

```
struct MemoryControlBlock {
    std::size_t       size_;
    MemoryControlBlock *prev_;
    MemoryControlBlock *prevFree_;
```

```
        MemoryControlBlock *nextFree_;
};
```

Field size_ holds the number of bytes of the current block. The last bit of size_ is reserved as an indicator of being busy. So block size is always even, but it reduces sizeof(MemoryControlBlock) by one byte. The next block located linearly in memory right next the current one can be accessed by adding the current block size to its address. Fields prevFree_ and nextFree_ are pointers to previous and next MemoryControlBlocks in the linked list of free blocks. The data part is located at the sizeof(MemoryControlBlock) bytes to the right of the control block address.

*C. Split and Consume*

Each time the user of our library [9] requests another block of dynamic memory, we iterate through the linked list of free blocks and check if there is a block, which is large enough to hold the requested number of bytes. If there was such a block, we split it first: divide it into two blocks, where the size of the first one equals the requested number of bytes. A split can not be done if there is not enough space for the second block with non-zero data part. Then the block is extracted from the list of free blocks and marked as busy. Address of block data is returned to the user.

If there is no such block in the list, that would fit the requested number of bytes, then ask the OS for more memory. We use mmap(2) on Posix and VirtualAlloc on Microsoft Windows as syscalls for dynamic memory allocation. Each of them returns a pointer to the mapped area. Firstly, two MemoryControlBlocks are placed at the beginning and at the end. The first one has prev_ == nullptr, while the last block is busy and has zero size to denote the end of the linear memory area. Then we shift the first block to the right by prepending a random sized free block. Then the block is splitted and returned to the user.

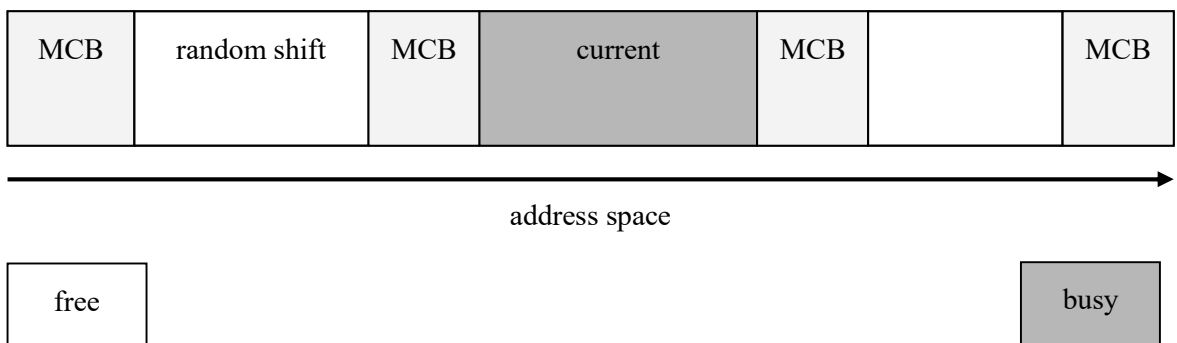

**Figure 1.** Memory layout of allocated chunk.

When a user frees memory by pointer, the corresponding MemoryControlBlock is sizeof(MemoryControlBlock) bytes to the left of the given pointer. After marking the block as free, we try to merge it with the adjacent blocks unless they are busy. So, if the left block is free, then it consumes the current one and the current becomes the left. Then do it again with the block to the right. Finally, the resulting block, which is being freed, might contain a non-zero number of memory pages, so we unmap them with munmap(2) or VirtualFree.

This technique allows us to reduce memory usage and time spent on iterating over the linked list of free blocks. However, this only increases the fragmentation of memory pages across the address space of a process and it may lead to inability to allocate another memory block or make it time inefficient at least. To avoid this problem we use a constant minimum continuous memory size. If the application allocates and deallocates objects very frequently and memory usage is not a bottleneck in the system, then we can raise the limit, so syscalls would be less frequent and memory pages would become less fragmented. That would make the application faster, since every syscall is done with processor interrupt. Unfortunately, this solution only applies to memory pages, not the objects within the application. If the program frequently allocates and deallocates small as well as large blocks of memory, then linear memory areas will be fragmented regardless of the aforementioned limit.

*D. Time and Memory Complexity*

Allocation takes $O(n)$ time in the worst case, where $n$ is the number of free blocks currently available: we need to traverse the linked list of free blocks in order to find one, which fits the requested number of bytes, or, if there is no such block, to ask the OS for more memory. In terms of space complexity, allocation is said to be constant: no additional information is accumulated during the iteration through the list of free blocks.

Deallocation takes constant both time and memory. There is only a limited number of steps in the algorithm that was described in section *C*.

*E. Memory Fragmentation*

Memory fragmentation is a phenomenon in which storage space is used inefficiently, reducing the capacity or performance and often both. The problem arises in runtime, when a program frequently allocates and deallocates blocks of memory of different sizes. The blocks, which are still busy, are scattered across the memory space of the process, thus making it difficult to find a block of a sufficient size to store the requested number of bytes. It increases list traversal time, frequency of system calls and has a negative impact on performance of the whole application. Moreover, it can lead to Out-of-Memory (OOM) errors, which are almost always fatal and cause the application to crash if it lacks OOM handling support [10].

There is an approach, which can be used to reduce memory fragmentation as well as make the allocation time constant: associate allocated chunks of memory pages with blocks of capacity within a given range and store multiple linked lists of free blocks with corresponding size. For example, there would be a separate list of small blocks with size in range $[0, 2^8)$, another list for blocks with size varying in range $[2^8, 2^{10})$, another one for $[2^{10}, 2^{12})$, etc. When the allocation function is called, the first block is taken from the list of blocks with size greater than or equal to the requested number of bytes. If the list is empty or does not exist yet, ask the OS for more memory, split it into blocks of corresponding size, add them to the list and take the first one. However, we *do not* use this approach since the key point here is that the block is not splitted, despite the requested number of bytes being less than its capacity, and it leads to inefficient memory packing with blocks and wasting space, which can be used to store blocks of smaller size.

*F. Split and Consume*

Different platforms have different requirements on data structures and allocated memory. Alignment is one of them and can significantly differ from system to system. For example, some 32-bit architectures require all *int* variables on a multiple of four. On some architectures, alignment requirements are absolute. On others, like x86, flouting them only comes with a performance penalty.

To ensure that the memory allocation performs time efficiently on different architectures, the allocator should be able to quickly tell whether a specific free MCB satisfies requested alignment or not. If we can shift this free MCB so that its position will be divisible by the proposed alignment, then we can be sure that the restriction is satisfied. The remaining part is to copy only MCB metadata without the need of copying the data it holds (because MCB is marked as free).

The computation of the lowest possible shift is a fast operation that requires simple division operations which can be evaluated even faster when the alignment equals to a power of two, which is the case on most systems. Therefore, the alignment allocation will be almost as efficient as the allocation without restrictions.

## 4. Discussion

Unfortunately, the algorithm is not thread-safe. When two or more threads try to allocate or deallocate a block, they both can modify the same linked list of free blocks by appending, removing, splitting or consuming one of the blocks, so it leads to simultaneous memory reads and writes, which, in turn, leads to an *undefined behaviour*.

The list of free blocks is shared across all threads and thread-local [11] storage duration would not be a solution here: the problem would arise again when, for example, free() is called in one thread on a pointer, which was allocated by another thread. The block being freed is merged with adjacent free blocks, which can happen to be located in the list of free blocks of another thread. So, there is no safe and efficient way to check if they were the first blocks in corresponding lists as it requires reading memory, which is managed by another thread. Moreover, these adjacent blocks may be currently used by another thread, while traversing its linked list of free blocks.

Nevertheless, thread-safety can be achieved by acquiring a lock on a whole allocator structure by each of allocation and deallocation operations, so that each thread has an exclusive access to it. But it was not a main concern of the current study, so we left it for future development.

## 5. Conclusion

In this paper we have presented an algorithm that is balanced in both requirements: efficiency and security. It can protect applications vulnerable to attacks like "use-after-free" while not being a bottleneck in terms of performance. Nevertheless, our implementation was not yet fully tested under some conditions, including multithreaded applications and streaming services, where memory fragmentation is more important and unexpected issues can arise.

Memory management is like every other algorithm: the speed can be increased, but it will cost more memory usage and vice versa. There is no perfect solution for all use cases - each solution is a certain balance between time and space complexity. The choice of dynamic memory management solution should depend on the application type and allocation frequency.